\documentclass[aps,prl,reprint,amsmath,amssymb,showpacs]{revtex4-1}
\usepackage{latexsym}
\usepackage{hyperref}
\usepackage{graphicx}

\begin{document}
\title{Omnidirectional transport in fully reconfigurable 2D optical ratchets}
\author{Alejandro \surname{V. Arzola}}
\email[]{alejandro@fisica.unam.mx}
\affiliation{Instituto de F\'isica, Universidad Nacional Aut\'onoma de M\'exico, Apdo. Postal 20-364, 01000 Cd. M\'exico, M\'exico}
\author{Petr \surname{J\'akl}}
\affiliation{Institute of Scientific Instruments of CAS, Kr\'alovopolsk\'a 147, 612 64, Brno, Czech Republic}
\author{Pavel \surname{Zem\'anek}}
\affiliation{Institute of Scientific Instruments of CAS, Kr\'alovopolsk\'a 147, 612 64, Brno, Czech Republic}
\author{Mario \surname{Villasante-Barahona}}
\affiliation{Instituto de F\'isica, Universidad Nacional Aut\'onoma de M\'exico, Apdo. Postal 20-364, 01000 Cd. M\'exico, M\'exico}
\author{Karen \surname{Volke-Sep\'ulveda}}
\affiliation{Instituto de F\'isica, Universidad Nacional Aut\'onoma de M\'exico, Apdo. Postal 20-364, 01000 Cd. M\'exico, M\'exico}
\begin{abstract} 
A fully reconfigurable two-dimensional (2D) rocking ratchet system created with holographic optical micromanipulation is presented. We can generate optical potentials with the geometry of any Bravais lattice in 2D and introduce a spatial asymmetry with arbitrary orientation. Nontrivial directed transport of Brownian particles along different directions is demonstrated numerically and experimentally, including on-axis, perpendicular and oblique with respect to an unbiased ac driving. The most important aspect to define the current direction is shown to be the asymmetry and not the driving orientation, and yet we show a system in which the asymmetry orientation of each potential well does not coincide with the transport direction, suggesting an additional symmetry breaking as a result of a coupling with the lattice configuration. Our experimental device, due to its versatility, opens up a new range of possibilities in the study of nonequilibrium dynamics at the microscopic level.
\end{abstract}
\maketitle
Initially motivated by the understanding of biological engines and the design of artificial nanodevices, the emergence of directed transport in the presence of unbiased external forces due to a spatiotemporal symmetry breaking has become a major research topic in different scientific areas \cite{hanggi_artificial_2009}. This intriguing phenomenon, known as ratchet effect, lies in the heart of nonequilibrium thermodynamics at microscopic scale. This model can explain the functioning of a number of systems in nature, such as molecular motors \cite{Hoffman_2016, vecchiarelli_propagating_2014, hu_directed_2015} or protein translocation processes \cite{gu_three_2010}. Additionally, schemes based on this mechanism have been implemented to sort biomolecules \cite{bader_dna_1999} and inorganic microparticles \cite{huang_continuous_2004}, to rectify the motion of cold atoms in optical lattices \cite{RGommers2006, Renzoni2008} and vortices in superconductors \cite{vorticesprl2005}, among others. The rich dynamics arising in ratchets becomes evident from the diverse phenomena that can be observed even in the simplest cases of one-dimensional (1D) systems, such as bidirectional transport depending on size, chaotic behavior and current reversals \cite{mateos2000chaotic, schreier_giant_1998, arzola2011experimental, malgaretti_running_2012, Arzola_PRE2013}. This is due to the delicate interplay among a whole set of parameters, encompassing the structure of a spatial potential, the modulation of an external driving, the strength of thermal noise and the properties of the particles. Studies on the influence of these aspects have paved the way to broaden our understanding of transport processes at the micro and nanoscale, but this area is far from complete.

Naturally, a degree of complexity and versatility is added in two-dimensional (2D) systems, which become very important in the context of electronic transport in 2D crystals like graphene \cite{drexler_magnetic_2013} and semiconductor artificial nanomaterials \cite{song_electron_2002}, for example. Among the studies of 2D ratchets, an explored path has been the use of symmetric spatial potentials either with a temporally asymmetric drive \cite{denisov_vortex_2008, sengupta_controlling_2004, Levedev_PhysRevA2009} or with an induced symmetry breaking due to a synchronization and phase coupling of two ac signals: a flashing potential and a symmetric rocking driving \cite{libal_reichhardt_2006, smith_2007}. Another common approach has been to tailor microfabriced substrates with posts or wells, introducing a space asymmetry \cite{SavelevPhysRevB2005, DerenyiPhysRevE1998, loutherback_deterministic_2009, nearfieldnano2016}. Motion rectification in this microstructured potentials, integrated in microfluidic schemes, as well as electrophoretic, superconductor, or solid-state devices, has been obtained, for example, for cells and bio-particles \cite{MortonLabChip2008, davis_deterministic_2006} and magnetic flux quanta in superconductors \cite{villegas_superconducting_2003}. In most of these devices, however, there is a static external force to induce transport along one direction, while the ratchet rectifies the transverse motion. There are few examples of ratchets having periodicity and dynamics in 2D involving only unbiased external (ac) forces to drive the system out of equilibrium to produce transport \cite{loutherback_deterministic_2009, nearfieldnano2016}.


In contrast with previous work, in this letter we present a fully reconfigurable 2D ratchet formed with a static asymmetric potential and an unbiased driving generating a rocking mechanism. Our experimental device, based on holographic optical micromanipulation, allows us to create any of the five Bravais lattices in 2D and introduce a spatial asymmetry of the individual potential wells along an arbitrary direction. We demonstrate, numerically and experimentally, controlled transport of Brownian particles in three different schemes of substrate potentials, where the ratcheting arises from the 2D nature of the potential. Motion rectification is obtained along the driving direction (on-axis current), along the transverse direction (lateral current), and also, for the first time to our knowledge, along an oblique direction, without any additional external force. Well beyond a novel micromanipulation device, our system constitutes an ideal experimental model to explore the physics of 2D ratchets. 

In the overdamped regime, the dynamics of a Brownian particle in a 2D rocking ratchet is described by:
\begin{equation}\label{eq_Langevin}
\dot{\vec{r}}(t) = -\frac{1}{\gamma} \nabla U\left(\vec{r}(t)\right) + \frac{1}{\gamma} \vec{F}(t)+\vec{\xi}(t),
\end{equation}
where $\vec{r}(t)=\left(x(t), y(t)\right)$, $\gamma$ is an effective drag coefficient, and $\vec{\xi}(t)$ represents the thermal noise, having a correlation function $\langle \xi_i(t) \xi_j(t')\rangle=2D \delta_{ij}\delta(t-t')$ with $i, j=x, y$. $D=k_B T/\gamma$ is the diffusion coefficient in the proximity of the surface \cite{supplementary}, different from that of a particle far from the surface $D_0$; $k_B$ is the Boltzman constant and $T$ the ambient temperature. The external driving is set along the $x$-direction, $\vec{F}(t)=(F(t), 0)$. The time modulation is given by \cite{arzola2011experimental, Arzola_PRE2013}: $F(t)=-\gamma v(t)$,  with $v(t+\tau)=v(t)$, where $v(t)=v_0$ if $0\leq t< \tau_1$; $v(t)=0$ if $\tau_1\leq t< \tau_1+\tau_0$; $v(t)=-v_0$ if $\tau_1+\tau_0\leq t< \tau-\tau_0$; and $v(t)=0$ if $\tau-\tau_0\leq t< \tau$; $v_0$ being a constant speed. The parameters $\tau_0$ and $\tau_1$, referred to as the waiting time and the activation time, respectively, satisfy $\tau=2(\tau_0+\tau_1)$.

\begin{figure}
\includegraphics[angle=0,width=3.2in]{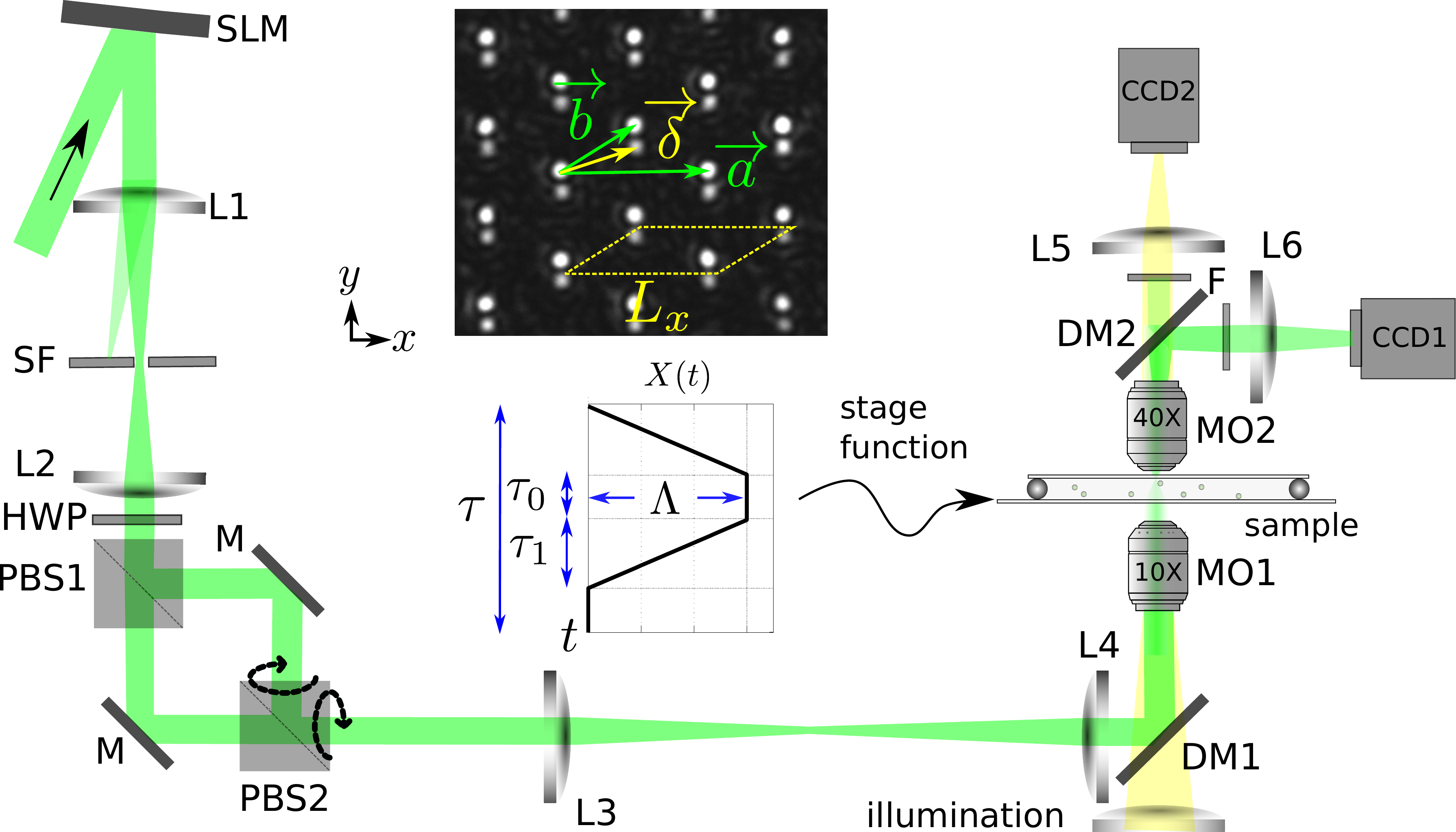}
\vspace{-0.3cm}
\caption{(Color online) Experimental setup: spatial light modulator (SLM), lenses (L), spatial filter (SF), half-wave plate (HWP), polarizing beam splitters (PBS), mirrors (M), dichroic mirrors (DM), microscope objectives (MO), attenuation filters (F) and cameras (CCD). The upper inset shows the primitive vectors of the main lattice $\vec{a}$ and $\vec{b}$, and the position of the clone lattice defined by $\vec{\delta}$. The inset in the middle illustrates the displacement of the piezo-stage $X(t)$.}\vspace{-0.8cm}\label{fig_setup}
\end{figure}

The potential is constituted by a collection of gaussian-shaped wells of widht $\sigma$ distributed over a lattice defined by the vectors $\vec{a}=(L_x, 0)$ and $\vec{b}=(b_x, b_y)$, illustrated in Fig. \ref{fig_setup}, and an identical clone-lattice of wells with a smaller depth by a factor $0<Q<1$, and which is shifted with respect to the main lattice to a position defined 
by $\vec{\delta}=(\delta_x, \delta_y)$. It can be expressed as:
\begin{equation}\label{eq_potencial}
U(\vec{r}) = -U_0\sum_{m,n}\left[u_{m,n}(x,y)+Q u_{m,n}(x-\delta_x,y-\delta_y)\right],
\end{equation}
where $u_{m,n}(x,y) = e^{-\frac{\left(\vec{r}-m\vec{a}-n\vec{b}\right)^2}{\sigma^2}}$. $U_0$ denotes the depth of 
each potential well of the main lattice, and the asymmetry of the potential depends on $Q$ and $\vec{\delta}$, which are related with experimental parameters as described in \cite{supplementary}.

The dynamics of the particle is characterized by the current $\vec{J}=\lim_{t\rightarrow \infty}\langle\vec{r}(t)\rangle/t$. However, if the waiting time $\tau_0$ is long enough, the particle can relax to an equilibrium position after each activation semi-cycle $\tau_1$ \cite{arzola2011experimental, Arzola_PRE2013}. Under these conditions, we define the dimensionless normalized current as $\vec{j}=(j_x, j_y)\approx\frac{1}{L_x}\frac{\vec{r}(q\tau)}{q}$, where $q=1, 2, 3,...$; this represents the number of periods the particle moves per unit cycle. The numerical calculations were done with the stochastic Runge-Kutta algorithm \cite{Rebecca_PRA1992}.

The experimental setup is illustrated in Fig.\ref{fig_setup}. A linearly polarized laser beam (532\,nm wavelength) is reflected on a spatial light modulator (SLM) displaying a computer generated hologram, which creates a pattern of multiple light spots of approximately equal intensity distributed over a 2D lattice. A Mach-Zehnder device splits the light beam into two orthogonally polarized components. Their relative intensities are controlled with a halfwave plate (HWP) before the polarizing beam splitter PBS1. The tilt of beamsplitter PBS2 is controlled with piezo actuators to shift the reflected pattern, corresponding to the clone, with respect to the transmitted one. Both co-propagating beams are projected onto the back aperture of a $10\times$ microscope objective (MO1). The dichroic mirror DM1 reflects the laser wavelength while letting pass the illumination of the imaging system. The desired intensity distribution forms at the back focal plane of MO1. With the imaging system we monitored the light pattern and the particles simultaneously.

We used monodisperse polystyrene microspheres with diameter $d=1.99 \mu$m immerse in water, with non-negligible Brownian motion at room temperature. The rocking mechanism is introduced by moving the sample cell sideways along the $x$ direction using a piezo-stage driven with velocity modulation $v(t)$, defined above. A plot of the displacement of the stage versus time, $X(t)$, is shown in Fig.\ref{fig_setup}. The activation time, $\tau_1$, and the amplitude of the periodic displacement, $\Lambda$, are the control parameters. Hence the maximum driving force on a particle is $|F_{max}|=\gamma v_0$, where $v_0=\Lambda/\tau_1$.

\begin{figure}
\includegraphics[angle=0,width=2.5in]{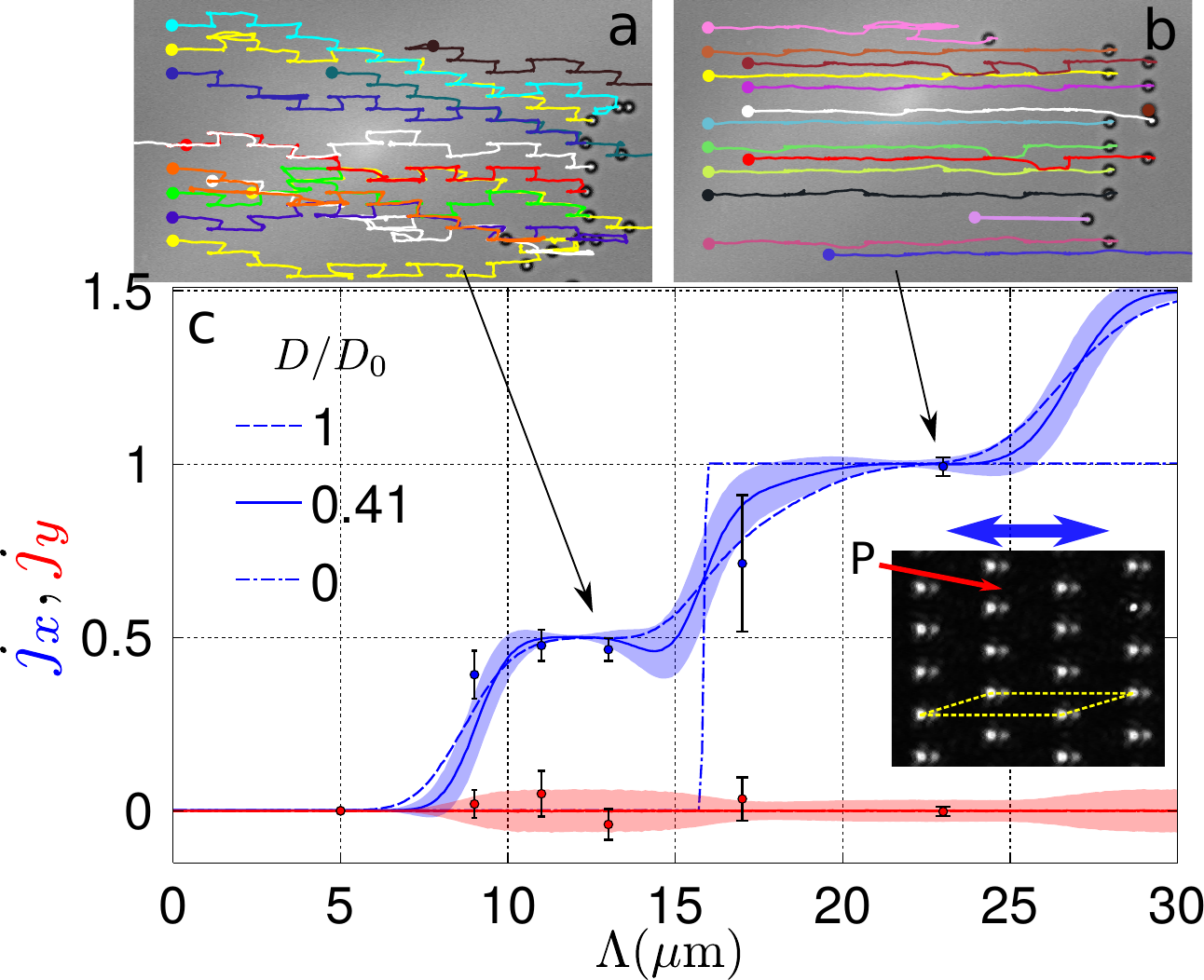}
\vspace{-0.3cm}
\caption{Experimental images showing the paths of an ensemble of particles for \cite{movie1}: (a) $\Lambda=13\,\mu$m and (b) $\Lambda=23\,\mu$m. (c) Numerical simulations for the components of the current, $j_x$  (blue) and $j_y$ (red), as a function of $\Lambda$, for three values of the relative diffusion coefficient $D/D_0$. The shadows around the curves for $D/D_0=0.41$ represent the standard deviations. The inset illustrates the intensity pattern and the direction of the driving (double arrow). Experimental results correspond to $D/D_0=0.41$ (markers). The parameters are: $L_x=14.8\,\mu$m, $b_x/L_x=0.5$, $b_y/L_x=0.15$, $v_0=21\,\mu m/$s.}\vspace{-0.8cm}
\label{fig_axial} 
\end{figure}

\begin{figure}
\includegraphics[angle=0,width=3.3in]{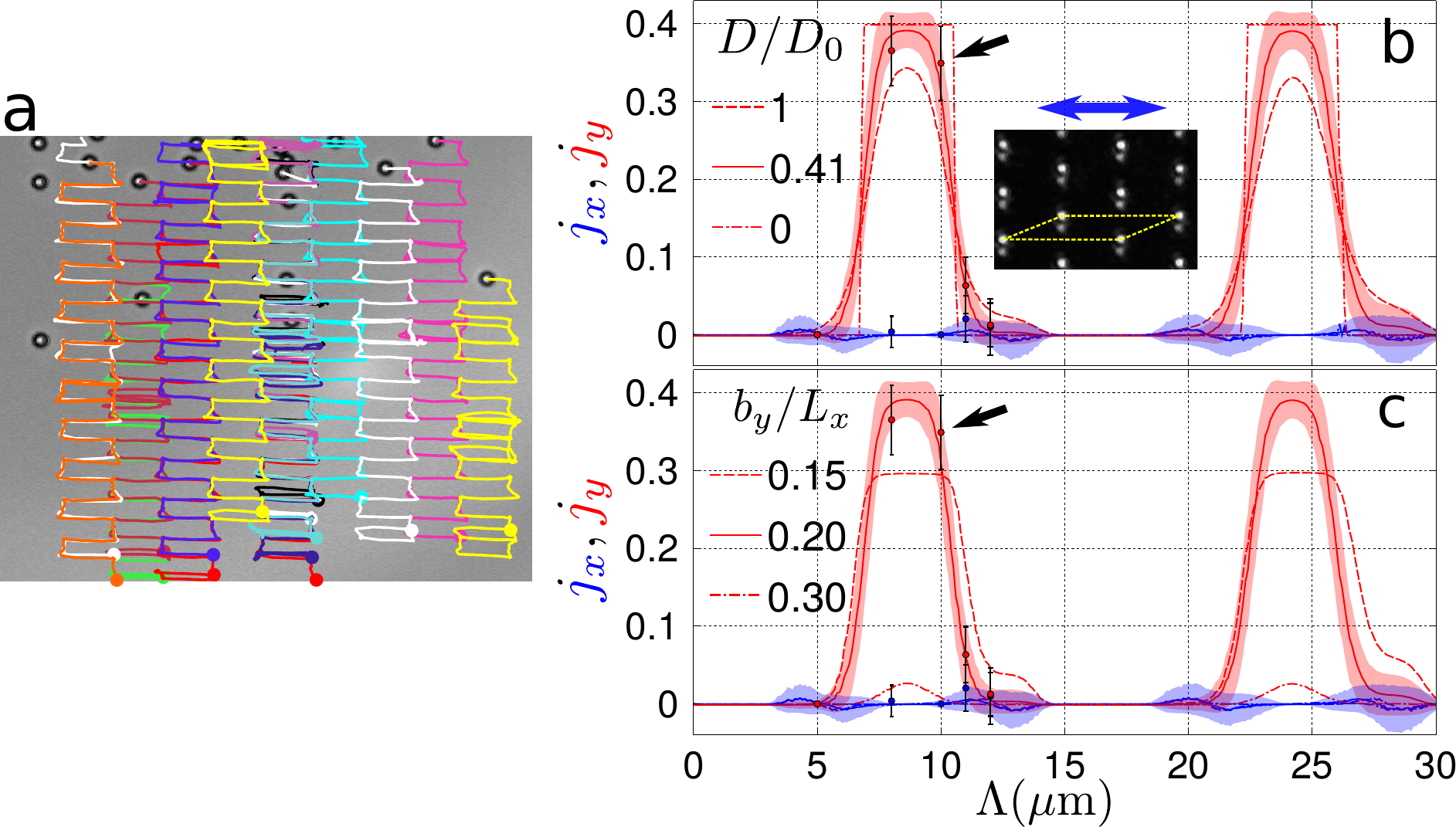}
\vspace{-0.3cm}
\caption{(a) Paths of an ensemble of particles for $\Lambda=10\,\mu $m. (b) Normalized current components ($j_x$  blue curves and $j_y$ red curves) \textit{vs} $\Lambda$ for the optical lattice shown in the inset, for three values of $D/D_0$. $L_x=14.8\,\mu \text{m}$, $v_0=50\,\mu \text{m}/$s, $b_x=0.5 Lx$, and $b_y=0.2Lx$. (c) Current components for $D/D_0=0.41$ and three values of $b_y/L_x$. The shadows in (b) and (c) correspond to the standard deviations for $D/D_0=0.41$ and $b_y/L_x=0.2$. The markers represent experimental values.}\vspace{-0.6cm}
\label{fig_lateral}
\end{figure}

\begin{figure}
\includegraphics[angle=0,width=2.5in]{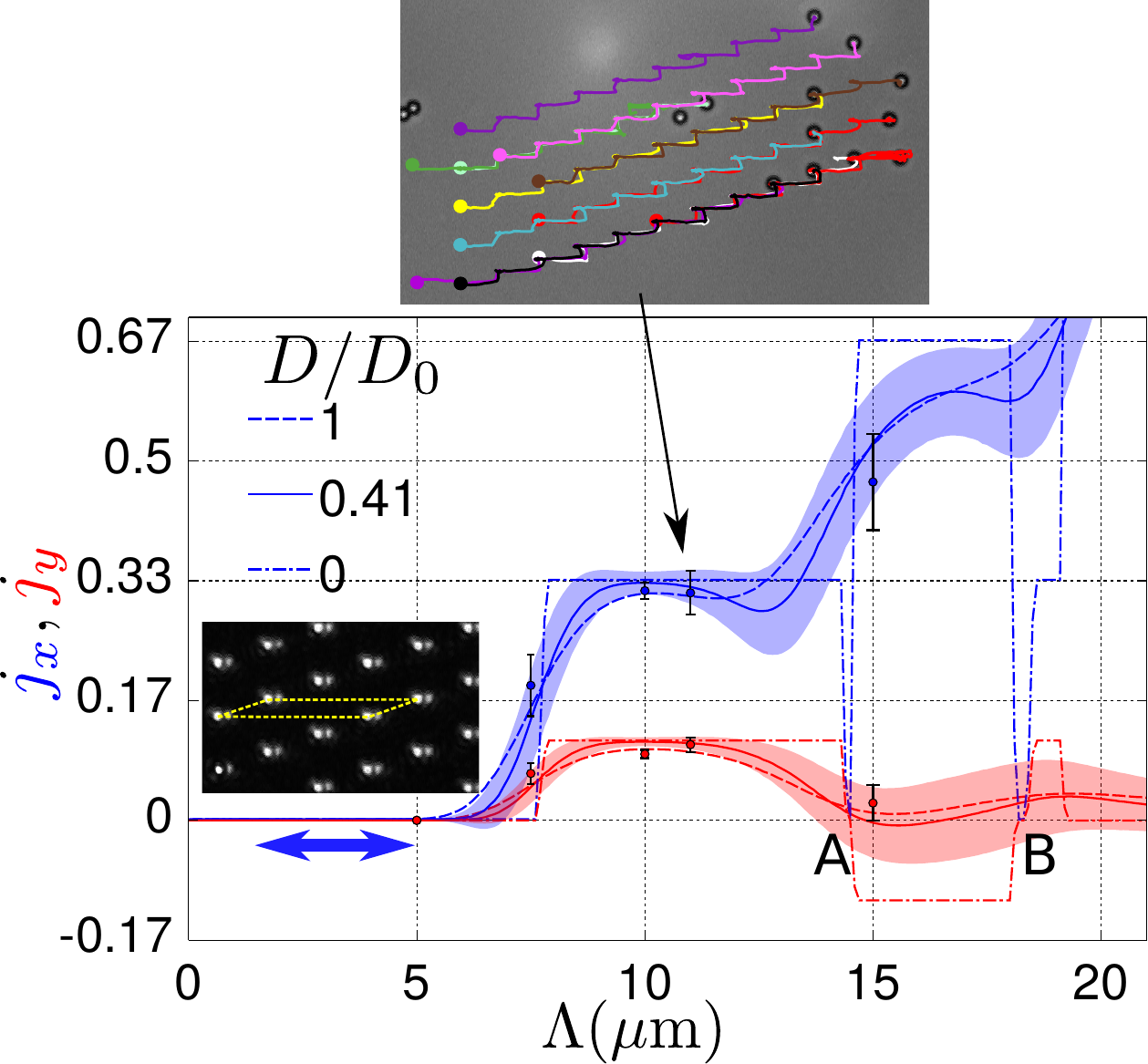}
\vspace{-0.3cm}
\caption{$j_x$ (blue) and $j_y$ (red) \textit{vs} $\Lambda$ for the optical lattice shown in the inset, for three values of $D/D_0$. $L_x=18.5\,\mu$m, $b_x= Lx/3$, $b_y=Lx/9$, and $v_0=24\,\mu \text{m}/$s. Markers denote experimental values for $D/D_0=0.41$. The image on top shows the trajectories of several particles for $\Lambda=11\,\mu $m, \cite{movie1}.}\vspace{-0.6cm}\label{fig_diagonal}
\end{figure}

We studied three schemes of 2D rocking ratchets. In all the experiments $\tau_0=2$s, and the values $Q=0.465\pm0.011$ and $|\vec{\delta}|=\delta=(1.42\pm0.05) \mu$m were chosen according to the calculations,  such that each potential well exhibits a maximum spatial asymmetry while its depth is enough to prevent diffusion of a trapped particle \cite{supplementary}. Therefore, particles mainly diffuse when they are in the intertraps space. In our simulations, we used an ensemble of 500 particles to determine $\vec{j}=(j_x, j_y)$ \textit{vs} $\Lambda$ for different values of the relative diffusion coefficient $D/D_0$, allowing us to compare the experimental results with the deterministic limit, $D\rightarrow 0$, and elucidate the role of noise strength. Experimentally, the positions of dozens of particles were tracked as a function of time from recorded videos, which are exemplified in \cite{movie1} for the cases illustrated in the images of Figs. \ref{fig_axial} to \ref{fig_diagonal}. The value of the activation time $\tau_1$ was adjusted as $\Lambda$ was varied, so as to keep $|F_{max}|$ constant for each set of experiments.

\textit{I. Axial current.} In the first scheme (Fig. \ref{fig_axial}), we have a centered rectangular lattice with a spatial asymmetry along $x$, described by: $\vec{a}/L_x=(1, 0)$ and $\vec{b}/L_x=(0.5, 0.15)$, with $L_x=14.8\,\mu$m, and $\vec{\delta}=(\delta, 0)$. The driving velocity is $v_0=21\,\mu m/$s; which allows the particles to escape from the wells in one direction (right), but not in the opposite (left). Figure \ref{fig_axial}c shows the numerical results for  $j_x$ (blue) and $j_y$ (red) \textit{vs} $\Lambda$ for $D/D_0=$0, 0.41, and 1. The shadows around the curves for $D/D_0=$0.41, which corresponds to our experiments \cite{supplementary}, represent the standard deviation of the simulation. We also plot experimental results (markers). Figures \ref{fig_axial}a-b illustrate the paths of several particles for two cases, departing from their initial position (colored circles on the left). 

Although the asymmetry of the potential and the driving are oriented along $x$, and therefore $j_y=0$, the dynamics of the system cannot, in general, be reduced to a 1D system due to diffusion. This is seen in Fig. \ref{fig_axial}a, which illustrates the role of diffusion and unstable equilibrium positions, such as the point P indicated in the inset of Fig. \ref{fig_axial}c. Namely, when the displacement of the particle is $\Delta x \sim L_x/2$, it ends up equally close to a pair of wells along $y$. Diffusion during the waiting time leads the bead to fall into one of these wells with equal probability. The result is a net current $j_x=1/2$, while the ensemble of particles spreads out laterally. In contrast, the current is null in the deterministic limit $D/D_0=$0 (dash-dotted curve) for this case. Even the tiniest thermal noise will kick the particle out from the unstable equilibrium position, onsetting the rectified motion. In Fig. \ref{fig_axial}b $\Delta x\simeq L_x$, therefore the particle reaches the next trap along $x$ with negligible role of diffusion, resulting in a practically deterministic 1D motion. Accordingly, when $D/D_0=0$, $j_x$ is a sharply stepped function with unitary step height, whereas for $D/D_0\neq0$ the curves exhibit twice the number of smoothed steps of height 1/2 starting at smaller values of $\Lambda$.

\textit{II. Lateral current.} In this case, we have a centered rectangular lattice but with a spatial asymmetry along $y$, so the particle is equally able to escape when the driving operates to the left or to the right. The parameters are: $\vec{a}/L_x=(1, 0)$, $\vec{b}/L_x=(0.5, 0.2)$, $L_x=14.8\,\mu$m, $\vec{\delta}=(0, \delta)$, and $v_0=50\,\mu \text{m}/$s. Figure \ref{fig_lateral}a illustrates the paths of several beads for $\Lambda = 0.68 L_x$, indicated with an arrow in Figs. \ref{fig_lateral}b-c. Blue and red curves correspond to $j_x$ and $j_y$, respectively, for three values of $D/D_0$. The current along $x$ is now negligible, while $j_y$ exhibits a periodic behavior. This is what we refer to as a lateral current, since it is perpendicular to the driving.

When the displacement of a particle in each activation semi-cycle is $\Delta x \sim L_x/2$, it gets attracted towards the potential well above it (along the $+y$ direction) during the waiting time, because of the well asymmetry. This process is repeated when the particle moves along $-x$ in the next activation semi-cycle. As a result, the bead moves a total distance of approximately $2b_y$ along $+y$ in each driving period (see the video in \cite{movie1}). As $\Lambda$ increases, such that $\Delta x \rightarrow L_x$ in each semi-cycle, $j_y\rightarrow 0$; in that case the particle only moves back and forth between two wells along $x$. The lateral current reaches a second maximum when $\Delta x\sim 3L_x/2$.

The magnitude of the lateral current can be optimized by varying the geometric parameter of the lattice $b_y/L_x$, as illustrated in Fig. \ref{fig_lateral}c for $D/D_0=0.41$. The closer the traps are to each other along the $y$ direction, for instance $b_y/L_x=0.15$ (dashed curve), the easier the particle falls into the top well, but the maximum lateral current is limited due to the small value of $b_y$. In contrast, for sufficiently large values of $b_y$, such as $b_y/L_x=0.3$ (dash-dotted curve), the probability for the particle to reach the top well diminishes. Larger values of $b_y$ lead to $j_y\rightarrow 0$. There is an optimum value of $b_y/L_x$, however, giving rise to a maximum lateral current. In our example this is close to $b_y/L_x=0.2$ (solid curve), although in general it depends on $D/D_0$ and the depth and shape of the potential wells. Although a lateral current has been observed in other ratchets operated by an ac microfluidic flow  \cite{loutherback_deterministic_2009}, the optimization would not be as straightforward as in our system, since we have fine control of the steps along $x$ and $y$.

\textit{III. Oblique current.} In the third scheme, we have an oblique lattice with a spatial asymmetry of each potential well along $x$, parallel to the rocking force, with parameters: $\vec{a}/L_x=(1, 0)$, $\vec{b}/L_x=(1/3, 1/9)$, $L_x=18.5\,\mu$m, and $\vec{\delta}=(\delta, 0)$. The driving velocity is $v_0=24\,\mu \text{m}/$s, which allows the particle to escape to the right but not to the left. Figure \ref{fig_diagonal} presents numerical results for $j_x$ (blue) and $j_y$ (red) for $D/D_0=0$, 0.41 and 1. The image on top shows an experiment for $\Lambda = 0.6 L_x$, indicated in the plot. Here the current of particles follows an oblique direction with respect to the forcing. To the best of our knowledge, this is the first time an oblique current is demonstrated in a 2D rachet in the absence of an external constant force. Remarkably, the nonzero current along $y$ suggests an additional symmetry breaking associated with a coupling between the asymmetry of each potential well and the geometry of the lattice itself. 

In this case, interesting features arise, particularly in the deterministic limit $D/D_0=0$ (dash-dotted curves). When $\Delta x\sim L_x/3$ to the right, the particle will be attracted by the nearest potential well along $+y$ during the waiting time, leading to $j_x=1/3$ and $j_y>0$. As $\Lambda$ increases, there is a value for which the particle reaches an unstable equilibrium position, equally close to a well above it and another below it. In the absence of noise, both components of the current become null (point A in Fig. \ref{fig_diagonal}). As $\Lambda$ keeps increasing so that $\Delta x \sim 2L_x/3$, the particle will be attracted to the well below it, therefore $j_x=2/3$ and $j_y<0$, implying a current reversal for $j_y$. There is a second unstable equilibrium position for a larger value of $\Lambda$ when the particle ends up equally close to the well below it and the one on its right (point B in Fig. \ref{fig_diagonal}). After this point, we can see a small region of the plots for which $j_y>0$ and $j_x=1/3$. This is because the particle is unable to reach a well when it moves forward, but on its way backwards it ends up close enough to the well above it at a distance $x \sim L_x/3$ from its initial position, falling in the well during the waiting time. Finally, when $\Lambda$ is such that $\Delta x \sim L_x$, then $j_x=1$ and $j_y=0$, hence the dynamics is reduced to a 1D motion along $x$. Notice that the effect of diffusion is to soften these sharply stepped curves \cite{Arzola_PRE2013}. Therefore, we expect the current reversal ($j_y<0$) to occur also for small values of $D/D_0$ and gradually disappear as diffusion increases.

In summary, we demonstrated directed transport of Brownian particles along different directions in a fully reconfigurable 2D optical ratchet potential activated with a rocking driving, showing in all cases a good agreement between experiment and simulations. The most important aspect to define the current direction was found to be the asymmetry and not the driving direction, and yet we showed a system in which the orientation of the asymmetry of each potential well did not coincide with the transport direction, suggesting an additional symmetry breaking as a result of a coupling with the lattice configuration. We also observed that the presence of unstable equilibrium positions gives rise to noise-induced transport. Our analysis unveiled only a few important aspects on the dynamics, but there is a plethora of effects that may arise as a function of different parameters, such as size-dependent interaction between particle and potential \cite{arzola2011experimental}, chaotic behavior \cite{mateos2000chaotic}, negative mobility \cite{magnasco1996}, etc. This illustrates how rich, versatile and powerful our experimental device is, openning up a whole new range of research possibilities in the study of 2D transport at the microscopic scale and colloidal systems as mechanical models for solid state physics, not only at the fundamental level, but also with relevant applications in the design of micro and nanodevices suitable for on-chip implementation.

\begin{acknowledgments}
This project was supported by CONACYT-CAS (171478), DGAPA-UNAM (PAPIIT-IA103615, PAPIIT-IN115614 and PAPIIT-IA104917), CSF (GB14-36681G), and MEYS CR (LO1212, LD 14069).
\end{acknowledgments}


\begin{thebibliography}{33}
\expandafter\ifx\csname natexlab\endcsname\relax\def\natexlab#1{#1}\fi
\expandafter\ifx\csname bibnamefont\endcsname\relax
  \def\bibnamefont#1{#1}\fi
\expandafter\ifx\csname bibfnamefont\endcsname\relax
  \def\bibfnamefont#1{#1}\fi
\expandafter\ifx\csname citenamefont\endcsname\relax
  \def\citenamefont#1{#1}\fi
\expandafter\ifx\csname url\endcsname\relax
  \def\url#1{\texttt{#1}}\fi
\expandafter\ifx\csname urlprefix\endcsname\relax\def\urlprefix{URL }\fi
\providecommand{\bibinfo}[2]{#2}
\providecommand{\eprint}[2][]{\url{#2}}

\bibitem[{\citenamefont{H\"anggi and
  Marchesoni}(2009)}]{hanggi_artificial_2009}
\bibinfo{author}{\bibfnamefont{P.}~\bibnamefont{H\"anggi}} \bibnamefont{and}
  \bibinfo{author}{\bibfnamefont{F.}~\bibnamefont{Marchesoni}},
  \bibinfo{journal}{Rev. Mod. Phys.} \textbf{\bibinfo{volume}{81}},
  \bibinfo{pages}{387} (\bibinfo{year}{2009}).

\bibitem[{\citenamefont{Hoffmann}(2016)}]{Hoffman_2016}
\bibinfo{author}{\bibfnamefont{P.~M.} \bibnamefont{Hoffmann}},
  \bibinfo{journal}{Rep. Prog. Phys.} \textbf{\bibinfo{volume}{79}},
  \bibinfo{pages}{032601} (\bibinfo{year}{2016}).

\bibitem[{\citenamefont{Vecchiarelli et~al.}(2014)\citenamefont{Vecchiarelli,
  Neuman, and Mizuuchi}}]{vecchiarelli_propagating_2014}
\bibinfo{author}{\bibfnamefont{A.~G.} \bibnamefont{Vecchiarelli}},
  \bibinfo{author}{\bibfnamefont{K.~C.} \bibnamefont{Neuman}},
  \bibnamefont{and} \bibinfo{author}{\bibfnamefont{K.}~\bibnamefont{Mizuuchi}},
  \bibinfo{journal}{Proceedings of the National Academy of Sciences}
  \textbf{\bibinfo{volume}{111}}, \bibinfo{pages}{4880} (\bibinfo{year}{2014}).

\bibitem[{\citenamefont{Hu et~al.}(2015)\citenamefont{Hu, Vecchiarelli,
  Mizuuchi, Neuman, and Liu}}]{hu_directed_2015}
\bibinfo{author}{\bibfnamefont{L.}~\bibnamefont{Hu}},
  \bibinfo{author}{\bibfnamefont{A.~G.} \bibnamefont{Vecchiarelli}},
  \bibinfo{author}{\bibfnamefont{K.}~\bibnamefont{Mizuuchi}},
  \bibinfo{author}{\bibfnamefont{K.~C.} \bibnamefont{Neuman}},
  \bibnamefont{and} \bibinfo{author}{\bibfnamefont{J.}~\bibnamefont{Liu}},
  \bibinfo{journal}{Proceedings of the National Academy of Sciences}
  \textbf{\bibinfo{volume}{112}}, \bibinfo{pages}{E7055}
  (\bibinfo{year}{2015}).

\bibitem[{\citenamefont{Gu and Rice}(2010)}]{gu_three_2010}
\bibinfo{author}{\bibfnamefont{M.}~\bibnamefont{Gu}} \bibnamefont{and}
  \bibinfo{author}{\bibfnamefont{C.~M.} \bibnamefont{Rice}},
  \bibinfo{journal}{Proceedings of the National Academy of Sciences}
  \textbf{\bibinfo{volume}{107}}, \bibinfo{pages}{521} (\bibinfo{year}{2010}).

\bibitem[{\citenamefont{Bader et~al.}(1999)\citenamefont{Bader, Hammond, Henck,
  Deem, McDermott, Bustillo, Simpson, Mulhern, and Rothberg}}]{bader_dna_1999}
\bibinfo{author}{\bibfnamefont{J.~S.} \bibnamefont{Bader}},
  \bibinfo{author}{\bibfnamefont{R.~W.} \bibnamefont{Hammond}},
  \bibinfo{author}{\bibfnamefont{S.~A.} \bibnamefont{Henck}},
  \bibinfo{author}{\bibfnamefont{M.~W.} \bibnamefont{Deem}},
  \bibinfo{author}{\bibfnamefont{G.~A.} \bibnamefont{McDermott}},
  \bibinfo{author}{\bibfnamefont{J.~M.} \bibnamefont{Bustillo}},
  \bibinfo{author}{\bibfnamefont{J.~W.} \bibnamefont{Simpson}},
  \bibinfo{author}{\bibfnamefont{G.~T.} \bibnamefont{Mulhern}},
  \bibnamefont{and} \bibinfo{author}{\bibfnamefont{J.~M.}
  \bibnamefont{Rothberg}}, \bibinfo{journal}{Proceedings of the National
  Academy of Sciences of the United States of America}
  \textbf{\bibinfo{volume}{96}}, \bibinfo{pages}{13165} (\bibinfo{year}{1999}).

\bibitem[{\citenamefont{Huang et~al.}(2004)\citenamefont{Huang, Cox, Austin,
  and Sturm}}]{huang_continuous_2004}
\bibinfo{author}{\bibfnamefont{L.~R.} \bibnamefont{Huang}},
  \bibinfo{author}{\bibfnamefont{E.~C.} \bibnamefont{Cox}},
  \bibinfo{author}{\bibfnamefont{R.~H.} \bibnamefont{Austin}},
  \bibnamefont{and} \bibinfo{author}{\bibfnamefont{J.~C.} \bibnamefont{Sturm}},
  \bibinfo{journal}{Science} \textbf{\bibinfo{volume}{304}},
  \bibinfo{pages}{987} (\bibinfo{year}{2004}).

\bibitem[{\citenamefont{Gommers et~al.}(2006)\citenamefont{Gommers, Denisov,
  and Renzoni}}]{RGommers2006}
\bibinfo{author}{\bibfnamefont{R.}~\bibnamefont{Gommers}},
  \bibinfo{author}{\bibfnamefont{S.}~\bibnamefont{Denisov}}, \bibnamefont{and}
  \bibinfo{author}{\bibfnamefont{F.}~\bibnamefont{Renzoni}},
  \bibinfo{journal}{Phys. Rev. Lett.} \textbf{\bibinfo{volume}{96}},
  \bibinfo{pages}{240604} (\bibinfo{year}{2006}).

\bibitem[{\citenamefont{Gommers et~al.}(2008)\citenamefont{Gommers, Lebedev,
  Brown, and Renzoni}}]{Renzoni2008}
\bibinfo{author}{\bibfnamefont{R.}~\bibnamefont{Gommers}},
  \bibinfo{author}{\bibfnamefont{V.}~\bibnamefont{Lebedev}},
  \bibinfo{author}{\bibfnamefont{M.}~\bibnamefont{Brown}}, \bibnamefont{and}
  \bibinfo{author}{\bibfnamefont{F.}~\bibnamefont{Renzoni}},
  \bibinfo{journal}{Phys. Rev. Lett.} \textbf{\bibinfo{volume}{100}},
  \bibinfo{pages}{040603} (\bibinfo{year}{2008}).

\bibitem[{\citenamefont{Togawa et~al.}(2005)\citenamefont{Togawa, Harada,
  Akashi, Kasai, Matsuda, Nori, Maeda, and Tonomura}}]{vorticesprl2005}
\bibinfo{author}{\bibfnamefont{Y.}~\bibnamefont{Togawa}},
  \bibinfo{author}{\bibfnamefont{K.}~\bibnamefont{Harada}},
  \bibinfo{author}{\bibfnamefont{T.}~\bibnamefont{Akashi}},
  \bibinfo{author}{\bibfnamefont{H.}~\bibnamefont{Kasai}},
  \bibinfo{author}{\bibfnamefont{T.}~\bibnamefont{Matsuda}},
  \bibinfo{author}{\bibfnamefont{F.}~\bibnamefont{Nori}},
  \bibinfo{author}{\bibfnamefont{A.}~\bibnamefont{Maeda}}, \bibnamefont{and}
  \bibinfo{author}{\bibfnamefont{A.}~\bibnamefont{Tonomura}},
  \bibinfo{journal}{Phys. Rev. Lett.} \textbf{\bibinfo{volume}{95}},
  \bibinfo{pages}{087002} (\bibinfo{year}{2005}).

\bibitem[{\citenamefont{Mateos}(2000)}]{mateos2000chaotic}
\bibinfo{author}{\bibfnamefont{J.}~\bibnamefont{Mateos}},
  \bibinfo{journal}{Phys. Rev. Lett.} \textbf{\bibinfo{volume}{84}},
  \bibinfo{pages}{258} (\bibinfo{year}{2000}).

\bibitem[{\citenamefont{Schreier et~al.}(1998)\citenamefont{Schreier, Reimann,
  H\"anggi, and Pollak}}]{schreier_giant_1998}
\bibinfo{author}{\bibfnamefont{M.}~\bibnamefont{Schreier}},
  \bibinfo{author}{\bibfnamefont{P.}~\bibnamefont{Reimann}},
  \bibinfo{author}{\bibfnamefont{P.}~\bibnamefont{H\"anggi}}, \bibnamefont{and}
  \bibinfo{author}{\bibfnamefont{E.}~\bibnamefont{Pollak}},
  \bibinfo{journal}{Europhys. Lett.} \textbf{\bibinfo{volume}{44}},
  \bibinfo{pages}{416} (\bibinfo{year}{1998}).

\bibitem[{\citenamefont{Arzola et~al.}(2011)\citenamefont{Arzola,
  Volke-Sep\'ulveda, and Mateos}}]{arzola2011experimental}
\bibinfo{author}{\bibfnamefont{A.~V.} \bibnamefont{Arzola}},
  \bibinfo{author}{\bibfnamefont{K.}~\bibnamefont{Volke-Sep\'ulveda}},
  \bibnamefont{and} \bibinfo{author}{\bibfnamefont{J.~L.}
  \bibnamefont{Mateos}}, \bibinfo{journal}{Phys. Rev. Lett.}
  \textbf{\bibinfo{volume}{106}}, \bibinfo{pages}{168104}
  (\bibinfo{year}{2011}).

\bibitem[{\citenamefont{Malgaretti et~al.}(2012)\citenamefont{Malgaretti,
  Pagonabarraga, and Frenkel}}]{malgaretti_running_2012}
\bibinfo{author}{\bibfnamefont{P.}~\bibnamefont{Malgaretti}},
  \bibinfo{author}{\bibfnamefont{I.}~\bibnamefont{Pagonabarraga}},
  \bibnamefont{and} \bibinfo{author}{\bibfnamefont{D.}~\bibnamefont{Frenkel}},
  \bibinfo{journal}{Phys. Rev. Lett.} \textbf{\bibinfo{volume}{109}}
  (\bibinfo{year}{2012}).

\bibitem[{\citenamefont{Arzola et~al.}(2013)\citenamefont{Arzola,
  Volke-Sep\'ulveda, and Mateos}}]{Arzola_PRE2013}
\bibinfo{author}{\bibfnamefont{A.~V.} \bibnamefont{Arzola}},
  \bibinfo{author}{\bibfnamefont{K.}~\bibnamefont{Volke-Sep\'ulveda}},
  \bibnamefont{and} \bibinfo{author}{\bibfnamefont{J.~L.}
  \bibnamefont{Mateos}}, \bibinfo{journal}{Phys. Rev. E}
  \textbf{\bibinfo{volume}{87}}, \bibinfo{pages}{062910}
  (\bibinfo{year}{2013}).

\bibitem[{\citenamefont{Drexler et~al.}(2013)\citenamefont{Drexler, Tarasenko,
  Olbrich, Karch, Hirmer, Müller, Gmitra, Fabian, Yakimova, Lara-Avila
  et~al.}}]{drexler_magnetic_2013}
\bibinfo{author}{\bibfnamefont{C.}~\bibnamefont{Drexler}},
  \bibinfo{author}{\bibfnamefont{S.~A.} \bibnamefont{Tarasenko}},
  \bibinfo{author}{\bibfnamefont{P.}~\bibnamefont{Olbrich}},
  \bibinfo{author}{\bibfnamefont{J.}~\bibnamefont{Karch}},
  \bibinfo{author}{\bibfnamefont{M.}~\bibnamefont{Hirmer}},
  \bibinfo{author}{\bibfnamefont{F.}~\bibnamefont{Müller}},
  \bibinfo{author}{\bibfnamefont{M.}~\bibnamefont{Gmitra}},
  \bibinfo{author}{\bibfnamefont{J.}~\bibnamefont{Fabian}},
  \bibinfo{author}{\bibfnamefont{R.}~\bibnamefont{Yakimova}},
  \bibinfo{author}{\bibfnamefont{S.}~\bibnamefont{Lara-Avila}},
  \bibnamefont{et~al.}, \bibinfo{journal}{Nature Nanotechnology}
  \textbf{\bibinfo{volume}{8}}, \bibinfo{pages}{104} (\bibinfo{year}{2013}).

\bibitem[{\citenamefont{Song}(2002)}]{song_electron_2002}
\bibinfo{author}{\bibfnamefont{A.}~\bibnamefont{Song}}, \bibinfo{journal}{Appl.
  Phys. A} \textbf{\bibinfo{volume}{75}}, \bibinfo{pages}{229}
  (\bibinfo{year}{2002}).

\bibitem[{\citenamefont{Denisov et~al.}(2008)\citenamefont{Denisov, Zolotaryuk,
  Flach, and Yevtushenko}}]{denisov_vortex_2008}
\bibinfo{author}{\bibfnamefont{S.}~\bibnamefont{Denisov}},
  \bibinfo{author}{\bibfnamefont{Y.}~\bibnamefont{Zolotaryuk}},
  \bibinfo{author}{\bibfnamefont{S.}~\bibnamefont{Flach}}, \bibnamefont{and}
  \bibinfo{author}{\bibfnamefont{O.}~\bibnamefont{Yevtushenko}},
  \bibinfo{journal}{Phys. Rev. Lett.} \textbf{\bibinfo{volume}{100}},
  \bibinfo{pages}{224102} (\bibinfo{year}{2008}).

\bibitem[{\citenamefont{Sengupta et~al.}(2004)\citenamefont{Sengupta, Guantes,
  Miret-Artés, and H\"anggi}}]{sengupta_controlling_2004}
\bibinfo{author}{\bibfnamefont{S.}~\bibnamefont{Sengupta}},
  \bibinfo{author}{\bibfnamefont{R.}~\bibnamefont{Guantes}},
  \bibinfo{author}{\bibfnamefont{S.}~\bibnamefont{Miret-Artés}},
  \bibnamefont{and} \bibinfo{author}{\bibfnamefont{P.}~\bibnamefont{H\"anggi}},
  \bibinfo{journal}{Physica A: Statistical Mechanics and its Applications}
  \textbf{\bibinfo{volume}{338}}, \bibinfo{pages}{406} (\bibinfo{year}{2004}).

\bibitem[{\citenamefont{Lebedev and Renzoni}(2009)}]{Levedev_PhysRevA2009}
\bibinfo{author}{\bibfnamefont{V.}~\bibnamefont{Lebedev}} \bibnamefont{and}
  \bibinfo{author}{\bibfnamefont{F.}~\bibnamefont{Renzoni}},
  \bibinfo{journal}{Phys. Rev. A} \textbf{\bibinfo{volume}{80}},
  \bibinfo{pages}{023422} (\bibinfo{year}{2009}).

\bibitem[{\citenamefont{Lib\'al et~al.}(2006)\citenamefont{Lib\'al, Reichhardt,
  Jank\'o, and Reichhardt}}]{libal_reichhardt_2006}
\bibinfo{author}{\bibfnamefont{A.}~\bibnamefont{Lib\'al}},
  \bibinfo{author}{\bibfnamefont{C.}~\bibnamefont{Reichhardt}},
  \bibinfo{author}{\bibfnamefont{B.}~\bibnamefont{Jank\'o}}, \bibnamefont{and}
  \bibinfo{author}{\bibfnamefont{C.~J.~O.} \bibnamefont{Reichhardt}},
  \bibinfo{journal}{Phys. Rev. Lett.} \textbf{\bibinfo{volume}{96}},
  \bibinfo{pages}{188301} (\bibinfo{year}{2006}).

\bibitem[{\citenamefont{Smith et~al.}(2007)\citenamefont{Smith, Spalding,
  Dholakia, and MacDonald}}]{smith_2007}
\bibinfo{author}{\bibfnamefont{R.~L.} \bibnamefont{Smith}},
  \bibinfo{author}{\bibfnamefont{G.~C.} \bibnamefont{Spalding}},
  \bibinfo{author}{\bibfnamefont{K.}~\bibnamefont{Dholakia}}, \bibnamefont{and}
  \bibinfo{author}{\bibfnamefont{M.~P.} \bibnamefont{MacDonald}},
  \bibinfo{journal}{Journal of Optics A: Pure and Applied Optics}
  \textbf{\bibinfo{volume}{9}}, \bibinfo{pages}{S134} (\bibinfo{year}{2007}).

\bibitem[{\citenamefont{Savel'ev et~al.}(2005)\citenamefont{Savel'ev, Misko,
  Marchesoni, and Nori}}]{SavelevPhysRevB2005}
\bibinfo{author}{\bibfnamefont{S.}~\bibnamefont{Savel'ev}},
  \bibinfo{author}{\bibfnamefont{V.}~\bibnamefont{Misko}},
  \bibinfo{author}{\bibfnamefont{F.}~\bibnamefont{Marchesoni}},
  \bibnamefont{and} \bibinfo{author}{\bibfnamefont{F.}~\bibnamefont{Nori}},
  \bibinfo{journal}{Phys. Rev. B} \textbf{\bibinfo{volume}{71}},
  \bibinfo{pages}{214303} (\bibinfo{year}{2005}).

\bibitem[{\citenamefont{Der\'enyi and
  Dean~Astumian}(1998)}]{DerenyiPhysRevE1998}
\bibinfo{author}{\bibfnamefont{I.}~\bibnamefont{Der\'enyi}} \bibnamefont{and}
  \bibinfo{author}{\bibfnamefont{R.}~\bibnamefont{Dean~Astumian}},
  \bibinfo{journal}{Phys. Rev. E} \textbf{\bibinfo{volume}{58}},
  \bibinfo{pages}{7781} (\bibinfo{year}{1998}).

\bibitem[{\citenamefont{Loutherback et~al.}(2009)\citenamefont{Loutherback,
  Puchalla, Austin, and Sturm}}]{loutherback_deterministic_2009}
\bibinfo{author}{\bibfnamefont{K.}~\bibnamefont{Loutherback}},
  \bibinfo{author}{\bibfnamefont{J.}~\bibnamefont{Puchalla}},
  \bibinfo{author}{\bibfnamefont{R.~H.} \bibnamefont{Austin}},
  \bibnamefont{and} \bibinfo{author}{\bibfnamefont{J.~C.} \bibnamefont{Sturm}},
  \bibinfo{journal}{Phys. Rev. Lett.} \textbf{\bibinfo{volume}{102}},
  \bibinfo{pages}{045301} (\bibinfo{year}{2009}).

\bibitem[{\citenamefont{Wu et~al.}(2016)\citenamefont{Wu, Huang, Jaquay, and
  Povinelli}}]{nearfieldnano2016}
\bibinfo{author}{\bibfnamefont{S.-H.} \bibnamefont{Wu}},
  \bibinfo{author}{\bibfnamefont{N.}~\bibnamefont{Huang}},
  \bibinfo{author}{\bibfnamefont{E.}~\bibnamefont{Jaquay}}, \bibnamefont{and}
  \bibinfo{author}{\bibfnamefont{M.~L.} \bibnamefont{Povinelli}},
  \bibinfo{journal}{Nano Letters} \textbf{\bibinfo{volume}{16}},
  \bibinfo{pages}{5261} (\bibinfo{year}{2016}).

\bibitem[{\citenamefont{Morton et~al.}(2008)\citenamefont{Morton, Loutherback,
  Inglis, Tsui, Sturm, Chou, and Austin}}]{MortonLabChip2008}
\bibinfo{author}{\bibfnamefont{K.~J.} \bibnamefont{Morton}},
  \bibinfo{author}{\bibfnamefont{K.}~\bibnamefont{Loutherback}},
  \bibinfo{author}{\bibfnamefont{D.~W.} \bibnamefont{Inglis}},
  \bibinfo{author}{\bibfnamefont{O.~K.} \bibnamefont{Tsui}},
  \bibinfo{author}{\bibfnamefont{J.~C.} \bibnamefont{Sturm}},
  \bibinfo{author}{\bibfnamefont{S.~Y.} \bibnamefont{Chou}}, \bibnamefont{and}
  \bibinfo{author}{\bibfnamefont{R.~H.} \bibnamefont{Austin}},
  \bibinfo{journal}{Lab Chip} \textbf{\bibinfo{volume}{8}},
  \bibinfo{pages}{1448} (\bibinfo{year}{2008}).

\bibitem[{\citenamefont{Davis et~al.}(2006)\citenamefont{Davis, Inglis, Morton,
  Lawrence, Huang, Chou, Sturm, and Austin}}]{davis_deterministic_2006}
\bibinfo{author}{\bibfnamefont{J.~A.} \bibnamefont{Davis}},
  \bibinfo{author}{\bibfnamefont{D.~W.} \bibnamefont{Inglis}},
  \bibinfo{author}{\bibfnamefont{K.~J.} \bibnamefont{Morton}},
  \bibinfo{author}{\bibfnamefont{D.~A.} \bibnamefont{Lawrence}},
  \bibinfo{author}{\bibfnamefont{L.~R.} \bibnamefont{Huang}},
  \bibinfo{author}{\bibfnamefont{S.~Y.} \bibnamefont{Chou}},
  \bibinfo{author}{\bibfnamefont{J.~C.} \bibnamefont{Sturm}}, \bibnamefont{and}
  \bibinfo{author}{\bibfnamefont{R.~H.} \bibnamefont{Austin}},
  \bibinfo{journal}{Proc. Natl. Acad. Sci. U S A}
  \textbf{\bibinfo{volume}{103}}, \bibinfo{pages}{14779}
  (\bibinfo{year}{2006}).

\bibitem[{\citenamefont{Villegas et~al.}(2003)\citenamefont{Villegas, Savel'ev,
  Nori, Gonzalez, Anguita, Garc{\'i}a, and
  Vicent}}]{villegas_superconducting_2003}
\bibinfo{author}{\bibfnamefont{J.~E.} \bibnamefont{Villegas}},
  \bibinfo{author}{\bibfnamefont{S.}~\bibnamefont{Savel'ev}},
  \bibinfo{author}{\bibfnamefont{F.}~\bibnamefont{Nori}},
  \bibinfo{author}{\bibfnamefont{E.~M.} \bibnamefont{Gonzalez}},
  \bibinfo{author}{\bibfnamefont{J.~V.} \bibnamefont{Anguita}},
  \bibinfo{author}{\bibfnamefont{R.}~\bibnamefont{Garc{\'i}a}},
  \bibnamefont{and} \bibinfo{author}{\bibfnamefont{J.~L.}
  \bibnamefont{Vicent}}, \bibinfo{journal}{Science}
  \textbf{\bibinfo{volume}{302}}, \bibinfo{pages}{1188} (\bibinfo{year}{2003}).

\bibitem[{sup()}]{supplementary}
\emph{\bibinfo{title}{See supplemental material at [url will be inserted by
  publisher] for a description of experimental details, which includes Refs.[34-35]}}.

\bibitem[{\citenamefont{Honeycutt}(1992)}]{Rebecca_PRA1992}
\bibinfo{author}{\bibfnamefont{R.~L.} \bibnamefont{Honeycutt}},
  \bibinfo{journal}{Phys. Rev. A} \textbf{\bibinfo{volume}{45}},
  \bibinfo{pages}{600} (\bibinfo{year}{1992}).

\bibitem[{mov()}]{movie1}
\emph{\bibinfo{title}{See supplemental material at [url will be inserted by
  publisher] for a video illustrating the reconfigurable ratchet potential and
  some examples of axial, lateral and oblique current. Video rate has been
  speeded up by a factor of 2.5.}}

\bibitem[{\citenamefont{Cecchi and Magnasco}(1996)}]{magnasco1996}
\bibinfo{author}{\bibfnamefont{G.~A.} \bibnamefont{Cecchi}} \bibnamefont{and}
  \bibinfo{author}{\bibfnamefont{M.~O.} \bibnamefont{Magnasco}},
  \bibinfo{journal}{Phys. Rev. Lett.} \textbf{\bibinfo{volume}{76}},
  \bibinfo{pages}{1968} (\bibinfo{year}{1996}).
  
\bibitem[{\citenamefont{Crocker and Grier}(1996)}]{Grier2}
\bibinfo{author}{\bibfnamefont{J.~C.} \bibnamefont{Crocker}} \bibnamefont{and}
  \bibinfo{author}{\bibfnamefont{D.~G.} \bibnamefont{Grier}},
  \bibinfo{journal}{Journal of Colloid and Interface Science}
  \textbf{\bibinfo{volume}{179}}, \bibinfo{pages}{298 } (\bibinfo{year}{1996}).

\bibitem[{\citenamefont{Franklin and Shattuck}(2015)}]{franklin_handbook_2015}
\bibinfo{editor}{\bibfnamefont{S.~V.} \bibnamefont{Franklin}} \bibnamefont{and}
  \bibinfo{editor}{\bibfnamefont{S.~M.} \bibnamefont{Shattuck}} eds.,
  \bibinfo{title}{Handbook of granular materials} (\bibinfo{publisher}{{CRC} Press, Taylor \& Francis Group},
  \bibinfo{address}{Boca Raton, {FL}}, 
  \bibinfo{year}{2015}).

\end{thebibliography}
\end{document}